# Fast Grid Splitting Detection for N-1 Contingency Analysis by Graph Computing

Yongli Zhu[1], Lingpeng Shi[2], Renchang Dai[1], Guangyi Liu[1]
[1]GEIRI North America, San Jose, USA (yongli.zhu@geirina.net, renchang.dai@geirina.net, guangyi.liu@geirina.net)
[2]Shanghai Electric Power Information & Communication Company, Shanghai, China (shilp@sh.sgcc.com.cn)

*Abstract*—In this study, a graph-computing based grid splitting detection algorithm is proposed for contingency analysis in a graph-based EMS (Energy Management System). The graph model of a power system is established by storing its bus-branch information into the corresponding vertex objects and edge objects of the graph database. Numerical comparison to an up-to-date serial computing algorithm is also investigated. Online tests on a real power system of China State Grid with 2752 buses and 3290 branches show that a 6 times speedup can be achieved, which lays a good foundation for advanced contingency analysis.

## I. INTRODUCTION

With the rapid development of the economy, the size and capacity of the power system have been increased. The system operators are confronted with challenges of handling a vast amount of information and making correct decisions in a faster time manner [1]. To prevent the system from disastrous events (e.g. disintegration of the whole power grid; loss of critical loads or generators, etc.) under any pre-considered contingency, the security assessment is crucial to the operation of large power systems [2][3]. One of the basic tasks of security assessment is the *N*-1 contingency analysis, which is used as the first step to discover any potential system fragilities. In practice, not all branches but only those whose loss will cause system overstressed need to be considered. This step is conventionally named as "fast screening", of which the output is a list of the most critical contingencies. The fast screening is typically achieved by some special branch flow updating algorithm [4], e.g. superposition method, DC power flow method, and sensitivity method by avoiding unnecessary LU factorizations of system admittance matrix after each contingency.

Depending on different needs and severity criteria, different screening scheme can be defined. Typically, contingencies which will cause *grid splitting* ("splitting cases") of the original power grid should be singled out and investigated separately. In different application scenarios, the handling methods can vary. For example, those splitting cases can be manually assigned some large severity indices to mark them as the "severest" in the final contingency list; or, a detailed power flow program can be performed on each isolated power islands, to check any violations of all the remaining branches in that island.

Thus, as an indispensable procedure for the fast screening, a speedy grid splitting detection algorithm will be critical for real-time monitoring and control of large power systems. In light of this, the parallel computing idea can be applied. Recently, power system applications based on graph database and graph computing have emerged such as power flow calculation [5] and electricity market applications [6]. Utilizing graph computing to speed up contingency analysis is still under exploration [7].

In this paper, a graph computing-based grid splitting detection algorithm for fast screening is developed on a commercial-level graph database, *TigerGraph* [8]. The proposed method is then tested on a real utility power grid for online application based on genuine operation data. Section II introduces the basic principle of the graph database and graph computing. Section III investigates different algorithms of grid splitting detection including the proposed graph computing algorithm. Section IV presents the case study on a real utility system of China State Grid. Conclusions and future directions are given in the last section.

## II. GRAPH DATABASE AND GRAPH COMPUTING

### A. Graph Database

Traditional *Relational Database* like MySQL is based on the relational model of data, and its basic operations like create, read, update, and delete (CRUD) are based on SQL (Structured Query Language). However, with the emerging of various new business models like social networks, which may have millions of "users" and their associated "relationships", the operation speed of the relational database will drop down quickly and become a bottleneck in the whole business.

Therefore, non-relational databases which utilize the intrinsic characteristics of real-world data to construct

This work is funded by State Grid Corporation Project SGSHXT00JFJS1700138

persistent storage with faster response, have been proposed, e.g. graph database.

## B. Graph Computing

Here, "graph computing" means "computing (either numerical or non-numerical) implemented on the graph database". The basic idea of graph computing can be summarized as three steps: "Communication", "Synchronization" and "Computation". After the original graph is divided and dispersed among different "Workers" (threads/processes/cores), the following procedures will happen within each "Super-step" (i.e. one high-level iteration):

- **Communicating** any necessary messages among different vertices and two consecutive "Super-steps"
- **Synchronization** of all intermediate results/messages from the vertices in the previous "Super-step"
- **Computation** is executed by all the "Workers" in current "Super-step" and executive orders are random.

Thus, graph computing provides an efficient parallel *programming paradigm* that is suitable for problems with natural graph structures like social networks and power grids.

## III. GRID SPLITTING DETECTION ALGORITHM

For the convenience of discussion on the graph algorithms, several useful concepts of the graph theory are recapped here:

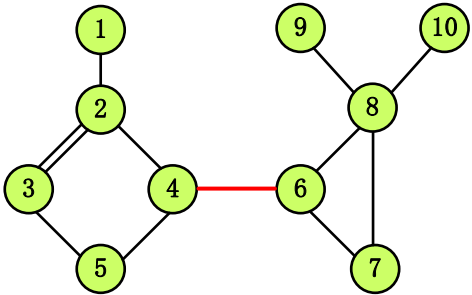

Figure 1. Example of a connected graph

Given an undirected graph, **G** = <**V**, **E**>, where **V** and **E** are the vertex set and edge set respectively. Denote the set cardinality as |**V**| and |**E**| respectively.

**Definition-1**: an undirected graph G is **connected**, if $\forall$ v, w $\in$ V, there exists a path in between them, e.g. Fig. 1.

**Definition-2**: a **connected component** of a given (undirected) graph is a subgraph in which any two vertices are connected to each other by at least one path but are not connected to any other vertices in the remaining graph. In Fig. 1, the graph itself is a connected component. If the bridge 4-6 is removed, then there will be two connected components: {1,2,3,4,5} and {6,7,8,9,10}.

**Definition-3**: a **bridge** is an edge when it is removed, the number of connected components will increase by one. Simply speaking, given a connected graph, the loss of a bridge will make the new graph unconnected. In Fig. 1, the edge 4-6 is a bridge.

Therefore, judging a power grid is split or not after removal of one edge is equivalent to *judging the edge is a bridge or not*. In graph theory language, this is equivalent to check that, the total number of the connected components in the new graph is *equal to one*.

Regarding the storage format for the undirected graph, the vertex is stored as an integer array whose elements are the vertex indices. As for the edges, there are two options:

- Adjacent matrix format: a two-dimension symmetric array $e$ is created, where $e[i][j]$ means there is an edge in between vertex-$i$ and vertex-$j$. For example, in Fig. 1, $e[6][7] = e[7][6] = 1$, $e[7][8] = e[8][7] = 1$.

- Adjacent linked list format: insert every target vertex in the adjacent linked list for vertex-$i$. For example, in Fig. 1, for vertex-7, its linked list may look like: List head of vertex-7 $\rightarrow$ vertex-8 $\rightarrow$ vertex-6.

## A. Special Handling for Parallel Lines

In real systems, there can be a lot of parallel lines (edges) in between certain two buses (vertices). Thus, the following techniques can be adopted: when reading the edge input data (suppose the *from* bus and the *to* bus are $i$ and $j$ respectively):

- For matrix format: increase $e[i][j]$ by one whenever a physical edge in between vertex-$i$ and vertex-$j$ is read. For example, $e[2][3] = 2$ and $e[3][2] = 2$.

- For the linked list format: insert the same neighbors in the linked list of vertex-$i$. For example, for vertex-2, its linked list may look like: List head of vertex-2 $\rightarrow$ vertex-3 $\rightarrow$ vertex-1 $\rightarrow$ vertex-3 $\rightarrow$ vertex-4.

In theory, any one of the parallel lines is not a bridge since removing it will not change the connectedness of the original graph. Thus, how to skip those parallel lines is the key point in programming. For matrix format, it is simple, i.e. only handling the case with $e[i][j] = 1$ but skip the other cases with $e[i][j] > 1$ (parallel lines exist) or $e[i][j] = 0$ (no edge exists). For the linked list format, the situation is slightly more complicated. The following approach is adopted:

- After reading all edge inputs, do a sorting (by ascending order of the vertex index) for each vertex's linked list, e.g. for vertex-2: List head of vertex-2 $\rightarrow$ vertex-1 $\rightarrow$ vertex-3 $\rightarrow$ vertex-3 $\rightarrow$ vertex-4.

- For the linked list of each vertex-$i$: if any consecutive list elements are the same, then skip all of them (i.e. moving the list iterator by the repetitive numbers)

## B. Serial Computing by DFS

The up-to-date serial computing algorithm in finding all the connected components for a given graph is the so-called **Tarjan algorithm**, invented by Robert E. Tarjan [9], which is based on Depth First Search (DFS). Its complexity is O(|**V**|+|**E**|) when the graph storage uses linked list format.

This algorithm introduces and maintains two record arrays: array "*low*" and array "*num*" during the graph traversal. The element of the array "*low*", i.e. *low*[$i$], is the "earliest reachable ancestor vertex index" for vertex-$i$. The element of array "*num*", say *num*[*i*], denotes the sequential "order" (i.e. the *recursion depth* of the DFS function) of each vertex when it is initially visited during the DFS traversal. For example, suppose edge 4-6 is removed, then after the overall DFS traversal terminates, the previous graph becomes to Fig. 2 (in a rotation view) and the final values of the two arrays are presented. It is observed that there are two DFS search trees: {4,5,3,2,1} and {6,7,8,10,9}. The changing process of the *low* array during the backtracking process is shown in Table I.

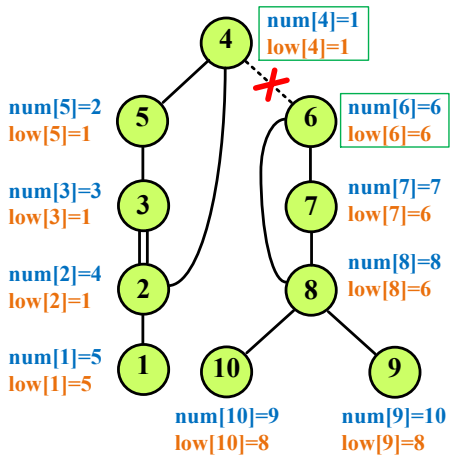

Figure 2. An illustration of the Tarjan algorithm

TABLE I. THE CHANGING PROCESS OF THE *LOW* ARRAY

|     | 1 | 2 | 3 | **4** | 5 | **6** | 7 | 8 | 9 | 10 |
|-----|---|---|---|---|---|---|---|---|---|----|
| *num* | 5 | 4 | 3 | 1 | 2 | 6 | 7 | 8 | 10 | 9 |
| *low* | 5 | – | – | – | – | – | – | – | 10 | – |
|     |   |   |   | ⋮ |   |   |   | ⋮ |   |    |
| *low* | 5 | 1 | 1 | **1** | 1 | **6** | 6 | 6 | 8 | 8 |

For current vertex-*i*, after all its unvisited neighbors have been traversed by DFS, its "*num*" value and "*low*" value will be checked to see if they are equal; if yes, then vertex-*i* is the root of the current DFS search tree and one new connected component is found.

In the last row of Table I, *low*[4] = *num*[4] = 1 and *low*[6] = *num*[6] = 6. Thus, there are totally two connected components. So, edge 4-6 is a bridge.

The C language-style pseudocode of the Tarjan algorithm in determining a given edge is a bridge or not is listed below. *CC_count* is the total connected components found in each case when a non-parallel edge is removed; *level* is the recursion depth of the DFS function. *Bridges_count* is the total number of found bridges.

### C. Graph Computing by BFS

In serial computing, BFS (Breadth First Search) can be also adopted but it needs to dynamically maintain a "queue" data structure with extra memory space. Thus, compared to DFS, it is relatively less used for bridge finding applications on serial computing platforms. However, in parallel computing, BFS can work to its advantage. It is naturally suitable for parallel implementation of the bridge finding algorithm since:

- BFS traverses the graph by "level" (breadth). This feature is in accordance with the communication mechanism of the graph database. So, it is easy for the graph computing platform to divide the traversal task.

- The extra space needed by BFS can be implicitly provided by the graph computing platform. Because by its design principle, the graph database itself will assign extra memory space to each "vertex" object whatever the task is BFS or DFS.

- Though it is possible to use DFS in parallel computing for bridge finding, the cost is the requirement of some sophisticated auxiliary algorithm plus specially designed communication mechanism of memory [10].

---

**Algorithm 1: Finding bridges by Tarjan algorithm**

1 **Input**: **G** = <**V**, **E** >, number of vertices *n*
2 **for** every edge *e* ∈ **E**
3     **if** (*e* is a parallel line)
4        continue     // skip and go to next different edge
5     Remove edge *e*
6     *CC_count* = 0, *level* = 0
7     *low*[*i*] ← 0, *num*[*i*] ← 0  ∀*i*
8     **for** every vertex-*i*
9        **if** (*num*[*i*]==0)  //if vertex-*i* is not visited
10          DFS_Tarjan (vertex-*i*, *CC_count*)
12     Recover edge *e*
13     **if** (*CC_count* > 1)  // find a bridge
14        *Bridges_count*++
15        Print *e* to file or mark *e* as a bridge
16 **End**
17 **Output**: an output file of all the bridges

---

**Algorithm 2: The DFS sub-function for Tarjan algorithm**

1 **void DFS_Tarjan** (vertex-*i*, *CC_count*)
2     *level*++
3     *num*[*i*] = *level*
4     *low*[*i*] =*level*
5     **for** every neighbor *j* of vertex-*i*
6        **if** (*num*[*j*]==0)  //if vertex-*j* is not visited
7          DFS_Tarjan (vertex-*j*, CC_count)
8          *low*[*i*] = min(*low*[*i*], *low*[*j*])
9        **else**
10          *low*[*i*] = min(*low*[*i*], *num*[*j*])
11     **if** (*low*[*i*] == *num*[*i*])  // a component has been found
12        *CC_count*++
13 **End**

---

The pseudocode on the graph computing platform for finding all the bridges and the corresponding connected components via BFS is shown as follows, where "**Parfor**" and "**ParWhile**" mean that the "for/while-loop" computation will be *implicitly* dispersed to parallel "Workers" and executed for all queried vertices" and/or "edges". In the graph database, each computation task is coded in a specific *query* similar to the *function* concept in serial programming.

---

**Algorithm 3: Main query of finding bridges**

---
1 **Input**: **G** = <**V**, **E** >, number of vertices *n*
2 **DummySet** = select *v* from **V** where (*e* is parallel line)
3 **Parfor** every edge *e* = (*v*, *w*) ∈ **E**
4   **if** (*e* is a parallel line)
5     continue  // skip and go to next different edge
6   *CC_count* = **BFS_CC** (*v*, *w*)
7   **if** (*CC_count* > 1)  // find a bridge
8     *Bridges_count*++
9     Print *e* to file or mark *e* as a bridge
10 **End**
11 **Output**: an output file of all the bridges

---

**Algorithm 4: The BFS sub-query for judging connectedness**

---
1 **int BFS_CC** (*u*, *v*)
2 **P** = {*u*, *v*}  // Parents set initialization
3 **C** = **select** *w* **from** all *e* = (*p*, *w*) **where** *p*∈**P**, *w*∉**P**
4   **if** (*p* == *u*)  //*w* is the neighbor of *u*
5     $CC1[w] = 1$
6   **if** (*p* == *v*)  //*w* is the neighbor of *v*
7     $CC2[w] = 1$
8 **Parfor** each *w*∈**C**
9   **if** ($CC1[w]$ == 1 and $CC2[w]$ == 1)
10     *CC_count* = 1  // *e* = (*u*, *v*) is not a bridge
11   **if** ($CC1[w]$ == 1 and $CC2[w]$ == 0)
12     *CC1_vts*++
13   **if** ($CC1[w]$ == 0 and $CC2[w]$ == 1)
14     *CC2_vts*++
15 **ParWhile** (*CC_count*==0 and *CC1_vts*>0 and *CC2_vts*>0)
16   *CC1_vts* = 0, *CC2_vts* = 0
17   **C** = **select** *w* **from** all *e* = (*c*, *w*)
18     **where** (*c*∈**C**) && ($CC1[w]$==0 || $CC2[w]$==0)
19       **if** ($CC1[c]$ == 1)  $CC1[w]$==1
20       **if** ($CC2[c]$ == 1)  $CC2[w]$==1
21   **Parfor** each *w*∈**C**
22     **if** ($CC1[w]$ == 1 and $CC2[w]$ == 1)
23       *CC_count* = 1  // *e* = (*u*, *v*) is not a bridge
24     **if** ($CC1[w]$ == 1 and $CC2[w]$ == 0)
25       *CC1_vts*++
26     **if** ($CC1[w]$ == 0 and $CC2[w]$ == 1)
27       *CC2_vts*++
28 **Return** (*CC_count* == 1) ? 1:2  //if not connected, return 2

---

In the above code, "CC1" ("CC2") is an indicator array to store the vertex that belongs to component-1 (2), i.e. inside the same component as the *from* (*to*) bus *u* (*v*) of the removed edge. "CC1_vts" ("CC2_vts") is the amount of found vertices belonging to the component-1 (2) during each BFS query. Fig. 3 illustrates one intermediate step of the above algorithm when edge 4-6 is removed.

## IV. CASE STUDY

In this section, a case study is performed on a real utility power grid. It contains 2752 buses and 3290 edges (including parallel lines). The original grid is connected as shown in Fig. 4. All the experiments here are implemented in C++ on a Linux server with 88 Intel Xeon CPU E5-2699A v4 @ 2.40GHz and 128GB memory.

### A. Graph computing v.s. Serial computing

For serial computing, both matrix and linked list formats are implemented for the Tarjan algorithm. Each test is performed five times and the average time cost is obtained. The final average time costs are shown in Table II. Compared to the best result by the serial algorithm (linked list), graph computing algorithm achieves about 6 times speedup.

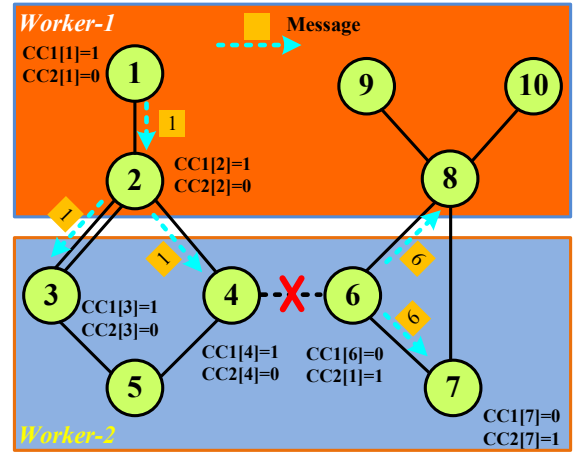

Figure 3. Illustration of Graph Computing based BFS for bridge finding

TABLE II. COMPUTATIONAL TIME FOR THE REAL POWER GRID

|  | Conventional method | Graph computing method |
|---|---|---|
| Adjacent matrix | 67820 ms | 205.54 ms |
| Adjacent linked list | 1350 ms | |

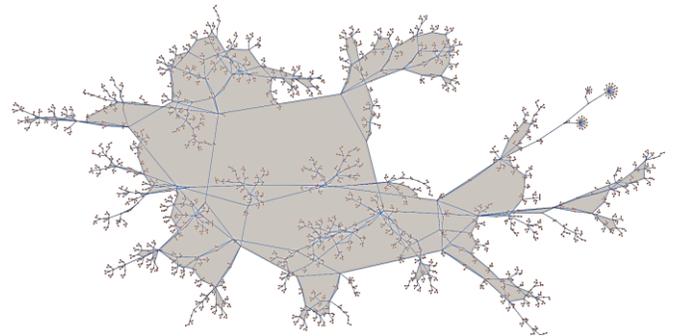

Figure 4. The single line diagram of a 2752-bus real power system

### B. Effect of different thread numbers

The time costs of the proposed graph computing algorithm by different thread numbers are shown in Table III and Fig. 5.

Thread numbers {1, 4, 8, 16, 32} are tested, and the time costs are recorded. As can be observed, 8 threads can achieve the best performance than others in our problem, which obtains a 5.75 times speedup than a single thread.

TABLE III. EFFECT OF DIFFERENT THREAD NUMBERS

| | Tread numbers | | | | |
|---|---|---|---|---|---|
| | 1 | 4 | 8 | 16 | 32 |
| Time (ms) | 1180.78 | 349.45 | **205.54** | 206.08 | 336.66 |

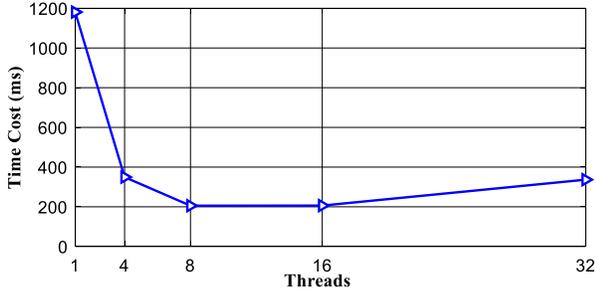

Figure 5. Time costs of graph computing under different thread numbers

### C. Application in Fast Screening of Contingencies

As an exemplary application of the proposed grid splitting detection algorithm, a fast screening case study is presented here. This fast screen program is also developed on the graph computing platform. In the screening scheme, only intact grid cases are considered, i.e. all the grid splitting cases will be ruled out based on the splitting detection results.

The DC power flow-based superposition method is adopted here for the fast screen calculation, which avoids the extra LU factorization for modified admittance matrix after an edge is removed. Its principle is concisely described as follows for an $n$-bus system. More details can be found in [4]:

$$\Delta \mathbf{P} = C \mathbf{M}_{pq} \boldsymbol{\theta}_0, \quad C = 1/(x_{ij} + \mathbf{M}_{pq} \mathbf{B}' \mathbf{M}_{pq}^T) \quad (1)$$

Where $\Delta \mathbf{P}=[\Delta P_i]$, $i=1...n$, is the bus power variation due to the removal of an edge; $x_{ij}$ is the branch reactance between bus-$i$ and $j$; $\mathbf{B}'$ is the pre-known nodal admittance matrix used for DC power flow in the base case (i.e. no edge is removed). $\mathbf{M}_{pq}$ is a row vector with "1" at the $p$-th place and "−1" at the $q$-th place: $\mathbf{M}_{pq} = [0, …, 1, …, −1, …,0]$. $\boldsymbol{\theta}_0$ is the pre-known bus angle vector in the base case. Then the following linear equations (2) can be solved (using the LU matrices of base case) for the angle variation $\Delta \boldsymbol{\theta} = [\Delta \theta_i]$, $i=1 ... n$.

$$\mathbf{B}' \Delta \boldsymbol{\theta} = \Delta \mathbf{P} \quad (2)$$

Finally, the branch flow between bus-$i$ and $j$ can be updated by (3), where $P_{ij,0}$ is the branch flow in the base case.

$$P_{ij} = P_{ij,0} + \Delta P_{ij}, \Delta P_{ij} = (\Delta \theta_i - \Delta \theta_j)/x_{ij} \quad (3)$$

To assess the importance of each (unsplit) $N$-1 case, the following *Severity Index* (*SI*) in (4) is defined and calculated, where $N_b$ is the total number of branches and $P_k$ is the active power flow of the $k$-th branch; $w_k$ is the pre-defined *weighting* constant for the $k$-th branch; $P_{k,\max}$ is the branch flow limit.

$$SI = \sum_{k=1}^{N_b} w_k \left( P_k / P_{k,\max} \right)^2 \quad (4)$$

After the grid splitting detection, 2256 "bridge" branches are detected and ruled out for fast screening. Then, all the low-voltage-level (<35kV) branches are further ruled out. Finally, 1137 branches are left for the fast screening analysis.

The top-5 cases of the fast screening results (with equal weights for all the non-bridge branches) are shown in Table IV. Case no. 2 and no. 3 have the same *from* and *to* buses but have different *SI* values because they are parallel lines with different parameters. The time cost is merely 128.589ms, which is fast enough for online applications.

TABLE IV. FAST SCREENING RESULTS BY GRAPH COMPUTING

| | Rankings of Fast Screening Results | | | | |
|---|---|---|---|---|---|
| | 1 | 2 | 3 | 4 | 5 |
| SI | 441.768 | 422.566 | 376.879 | 257.446 | 255.353 |
| Branch i-j | 2046-2460 | 1870-118 | 1870-118 | 289-1244 | 1108-759 |
| Time | 128.589 ms | | | | |

## V. CONCLUSIONS

In this paper, a graph computing-based grid splitting detection algorithm is implemented. Its application in fast screening is demonstrated. Results on the real system highlight its speed advantage. Some undergoing work includes 1) developing a detailed power flow program to validate the fast screening results and 2) considering complicated events like *N*-1-1 contingencies in our graph computing platform.